# Mixed charged pseudo-*zwitterionic* mesoporous silica nanoparticles with low-fouling and reduced cell uptake properties

Noemí Encinas[a], Mercedes Angulo[a], Carlos Astorga[a], Montserrat Colilla[a,b], Isabel Izquierdo-Barba[a,b*], María Vallet-Regí[a,b*]

[a] *Chemistry in Pharmaceutical Sciences Department, Inorganic and Bioinorganic Chemistry Unit, Universidad Complutense de Madrid. Instituto de Investigación Sanitaria Hospital 12 de Octubre i+12, Plaza Ramón y Cajal s/n, 28040 Madrid, Spain*

[b]*CIBER Bioengineering, Biomaterials and Nanomedicine, CIBER-BBN, Madrid, Spain*

**Abstract**

The design of drug delivery systems needs to consider biocompatibility and host body recognition for an adequate actuation. In this work, mesoporous silica nanoparticles (MSNs) surfaces were successfully modified with two silane molecules to provide mixed-charge brushes ($-NH_3^{\oplus}/-PO_3^{\ominus}$) and well evaluated in terms of surface properties, low-fouling capability and cell uptake in comparison to PEGylated MSNs. The modification process consists in the simultaneous direct-grafting of hydrolysable short chain amino (aminopropyl silanetriol, APST) and phosphonate-based (trihydroxy-silyl-propyl-methyl-phosphonate, THSPMP) silane molecules able to provide a pseudo-*zwitterionic* nature under physiological pH conditions. Results confirmed that both mixed-charged pseudo-*zwitterionic* MSNs (ZMSN) and PEG-MSN display a significant reduction of serum protein adhesion and macrophages uptake with respect to pristine MSNs. In the case of ZMSNs, his reduction is up to a 70-90% for protein adsorption and c.a. 60% for cellular uptake. This pseudo-*zwitterionic* modification has been focused on the aim of local treatment of bacterial infections through the synergistic effect between the capability of the nanocarriers to transport and deliver specific drugs and the electrostatic interactions between the mixed-charge systems and bacteria. These findings open promising future expectations for the effective treatment of bacterial infections through the use mixed-charge pseudo-*zwitterionic* MSNs furtive to macrophages and with antimicrobial properties.

**Introduction**

Currently, bone infection represents one of the most common complications in clinical practice with serious social and economic implications [1]. It is associated with significant morbidity for patients and substantial health care costs. A major issue to be considered is the increased multi-drug resistant (MDR) bacteria that is estimated to cause more than 10 million deaths via infection for 2050 [2-4]. Therefore, there is a great interest to promote new alternative therapies [5-10]. Mesoporous silica nanoparticles (MSNs) are highly attractive candidates for targeting drug delivery accounting for their special features and were first described for this purpose by Vallet-Regí in 2001 [11]. Some of these features are their high specific surface area, large and tunable pore size, drug loading capacity, simple synthesis, versatility for chemical modification and biocompatibility [12]. Consequently, MSNs could be an effective solution for infection treatment as it has been recently reported [13]. In this sense, MSNs offer the possibility of tackle MDR facilitating cellular accessibility (vectorization) and improving drug efficiency due to the ease of surface functionalization and high loading ability (therefore, it is possible to tune local drug administration avoiding dosage toxic side-effects by reducing the minimum inhibitory concentration compared to the free drug) [14]. However, it is known the tendency of these nanocarriers to aggregate and undergo nonspecific protein adsorption (formation of "protein corona") when encounter biological fluids. These mechanisms negatively affect their efficiency and clearance by the body immune system through macrophage fate [15-17]. Therefore, the design of nanoparticles with non-fouling properties against proteins is key to reduce their recognition by the immune system and thus optimize their applications in nanomedicine [18]. The most commonly used strategy to cloak MSNs and increase blood circulation time is grafting with hydrophilic polyethylene glycol (PEG) due to its ability to create a steric repulsion barrier against serum proteins through a water hydrogen-bound layer [19]. However, PEGylation can increase nanoparticle hydrodynamic radius and undergoes oxidation in the presence of oxygen and transition metal ions [20].

*Zwitterionic* nanoparticles raise a new challenge for the translation of inorganic nanoparticles into clinical practice. As described by Laschewsky and Rosenhahn and following the IUPAC definition, *zwitterionic* surfaces are a sub-class of polyampholytes which possess an equal number of both positive and negative charges on the same pendant group maintaining overall electrical neutrality [21]. The low-fouling capability of *zwitterionic* surfaces is related to the formation of a hydration layer more strongly compared to hydrophilic materials, forming a physical and energetic barrier that hinders nonspecific proteins adhesion [22,23]. Initial efforts were focused on the use of *zwitterionic* polymers bearing mixed positively and negatively charged moieties within the same chain with an overall charge neutrality [24,25]. In general, the main approaches to graft *zwitterionic* polymers to nanoparticles are: (i) the surface-initiated atom transfer radical polymerization (SI-ATRP), which has been applied with poly(sulfobetaine) (pSB) [26] and polycarboxybetaine (pCB) derivatives [27] and (ii) the surface reversible addition-fragmentation chain transfer (RAFT) applied mainly to graft SB copolymers to MSNs [28]. Furthermore, it is possible to confer *zwitterionic* nature by decorating their surface with low-molecular weight moieties bearing the same number of negative and positive charges. Although these methods usually require several synthetic steps involving different intermediate products, they offer diverse advantages compared to *zwitterionic* polymers since they are relatively more simple and lead to more biocompatible surfaces. For example, sulfobetaine derivatives [29] or amino acids such as cysteine and lysine can be used [30,31]. In this case, amino acids cannot be strictly considered as *zwitterions* due to the presence of ionizable groups, which provides a non-permanent (pH dependent) *zwitterion*-like behavior at the isoelectric point.

A substantial synthetic advance in the development of *zwitterionic*-like surfaces of biomaterials has been the use of more direct and simple grafting methods based on sol-gel chemistry with organosilanes [32-34]. In this case, taking advantage of the presence of hydroxyl (-OH)-containing biomaterials (i.e. silica) it is feasible to simultaneously attach two organosilanes exhibiting functional groups with compensated positive and negative charges. Bioceramic surfaces as SBA-15 and hydroxyapatite has been modified by this approach to confer surfaces not only of high resistance to nonspecific protein adsorption, but also to bacterial adhesion and/or bacterial biofilm [33-35]. This strategy allows for tailoring the *zwitterionic*-like properties of the materials by adjusting the molar ratio of the different reactants.

Herein, MSNs were designed and synthesized via simultaneous direct grafting of short chains of amino (aminopropyl silanetriol, APST) and phosphonate-based (trihydroxy-silyl-propyl-methyl-phosphonate, THSPMP) silane molecules. The main hint of our study is the ease and feasibility of the method, based on the combination of short-chain fully hydrolysable silane molecules capable to produce anionic/cationic charge separation on the MSNs surface under physiological conditions (pH=7.4), keeping an overall electrical neutrality. In this sense, a non-permanent and pH-dependent *zwitterionic*-like behaviour (hereafter named as pseudo-*zwitterionic*) was obtained. Antifouling ability through reduced protein adsorption of serum proteins was evaluated by gel electrophoresis (SDS-PAGE) experiments. The low cell uptake against RAW 264.7 macrophages was proved through laser scanning confocal microscopy and flow cytometry assays. For comparative purposes, PEGylated MSNs were fabricated using a similarly short chain polymer. This functionalization has been focused on the specific infection treatment. Furthermore, the capacity to encapsulate drugs to target bacterial infection was assessed through loading with a fluoroquinolone broad-spectrum antibiotic as levofloxacin (LEVO).

**Materials and Methods**

*1. Chemicals*

Fluorescently labelled MCM-41 MSNs were synthesized through a modified Stöber reaction [36] by using: tetraethylorthosilicate (TEOS, 98%), 2M sodium hydroxide 2M solution (> 97 %, NaOH), cetyltrimethylammonium bromide (> 98%, CTAB), aminopropyltriethoxysilane (> 97%, APTES), fluorescein isothiocyanate (> 99%, FITC), ammonium nitrate ($NH_4NO_3$) and ethanol, all of them purchased from Sigma-Aldrich and used as-received. Amino-phosphonate functionalization was performed by using aminopropyl silanetriol (APST, 96%) and trihydroxy-silyl-propyl-methyl-phosphonate (THSPMP) obtained from ABCR GmbH & Co. KG. PEGylated nanoparticles (PEG) were synthesized by functionalization with 3-[methoxy(polyethyleneoxy)propyl]trimethoxysilane] of 6-9 units (90%, Gelest) using anhydrous toluene (99.8%, Sigma-Aldrich). Solvents of high purity (> 99.5% 2-propanol, methanol, and toluene) were purchased from Sigma-Aldrich, as well as LEVO antibiotic. Deionized water was purified by passage through a Milli-Q Advantage A-10 purification system (Millipore Corporation) to a final resistivity of 18.2 MΩ.cm.

As-synthesized MSNs were properly characterized through the following methods: small-angle powder X-ray diffraction (XRD); $N_2$ adsorption porosimetry (BET); dynamic light scattering (DLS) and ζ-potential measurements; attenuated total reflectance Fourier transform infrared spectroscopy (ATR-FTIR); solution state NMR spectroscopy; high resolution magic angle spinning (HR-MAS) NMR spectroscopy; solid state MAS NMR and cross polarization (CP) MAS NMR spectroscopy; elemental chemical analysis; thermogravimetric and differential thermal analysis (TGA); transmission electron microscopy (TEM) and scanning electron microscopy (SEM). Description of devices and measurement conditions are listed in the Supporting Information.

*2. Synthesis of fluorescently labelled MCM-41 nanoparticles (MSNs)*

Firstly, the silylated labelling solution was prepared by adding 2.2 µL of APTES (0.01 mmol) to a mixture of 1 mg of FITC (0.0025 mmol) in 0.1 mL of TEOS and stirred at room temperature under dark conditions for 2 h. Afterwards 5 mL of TEOS (22.4 mmol) were added to the mixture and the resulting solution was kept in a syringe dispenser. The micelle templating solution was prepared by dissolving 1 g of CTAB (2.74 mmol) in 480 mL of water, adding 3.5 mL of NaOH 2 M and heating to 80 ºC upon vigorous stirring while keeping a stable vortex for 45 min. Then, the labelling solution (APTES-FITC-TEOS) was added at a constant 0.33 mL/min rate and kept stirring for 2 h at 80 ºC. The resulting colloidal suspension was cooled down in a water bath and particles were recovered through washing and centrifuging steps (water, ethanol x3). The surfactant template was removed by ion exchange with constant overnight stirring and reflux (80 ºC) of nanoparticles suspension in a $NH_4NO_3$ (10g/L):EtOH:$H_2O$ (95:5) solution (1 g nanoparticles/350 mL solution). After extraction the particles were washed through centrifugation with deionized water and EtOH (x2) and dried overnight under vacuum at 30 ºC [13].

*3. Synthesis of fluorescent mixed-charged amino-phosphonate pseudo-zwitterionic MCM-41 particles (ZMSNs)*

For covalent anchorage of cationic/anionic agents on the external surface of MSNs, the initial labelled nanoparticles containing the CTAB template (500 mg) were dried and degassed under vacuum overnight and then re-dispersed in 100 mL of absolute EtOH by ultrasound cycles (especial care was taken to avoid particle agglomeration) for 20 min under $N_2$ atmosphere. A mixture of the silane precursors (APST:THSPMP) was simultaneously added in varying molar ratios to optimize the pseudo-*zwitterionic* properties of the functionalized nanoparticles. This bifunctionalization has been modified from reference [31]. The reaction was kept overnight (inert atmosphere, constant stirring and reflux at 80 ºC) before thorough washing and recovery under centrifugation with EtOH (x2) and methanol followed by drying overnight under vacuum at 30 ºC. After outer surface functionalization, ZMSNs were subjected to surfactant extraction following the procedure descripted before.

*4. Synthesis of fluorescent PEGylated MCM-41 particles (PEGMSN)*

For PEGylation of the outer surface of MSNs, CTAB-filled labelled colloids were degassed overnight and re-suspended in 100 mL of anhydrous toluene under $N_2$ atmosphere. In this case the alcoxy PEG precursor was added (0.38 mmol/g) to the suspension and the reaction underwent overnight at 110 °C reflux and stirring. The product was isolated by centrifugation and subsequent washing with toluene, 2-propanol, ethanol and methanol. For further use the solid nanoparticles were dried overnight under vacuum at 30 ºC. After outer surface functionalization, PEGylated MSNs were subjected to surfactant extraction following the procedure descripted before.

*5. Drug loading and ''in vial'' release assays*

LEVO was loaded into the inner structure of the MSNs (pores) through impregnation method by soaking the dried nanoparticles in a 0.008 M solution of LEVO in absolute ethanol (5 mg nanoparticles/mL antibiotic stock solution) [37]. The colloidal suspension was stirred at room temperature for 16 h under dark conditions, followed by recovering of the product by filtration and drying under vacuum. The amount of LEVO loaded was determined by indirect method, calculating the total amount of LEVO released by fluorescence and compared with TGA and elemental chemical analyses results. Release kinetics of loaded LEVO was determined through fluorescence spectrometry (BiotekPowerwave XS spectrofluorimeter, version 1.00.14 of the Gen5 program, with $\lambda_{ex}$ = 292 nm and $\lambda_{em}$ = 494 nm) under physiological conditions (pH = 7.4, 37 ºC) in phosphate buffered solution (PBS) [37]. The system consisted on a double-chamber cuvette equipped with sample and analysis compartments separated by a dialysis membrane with a 12 kDa molecular weight cut-off membrane allowing selective LEVO diffusion. 168 µL of the nanoparticles loaded with LEVO (MSN@LEVO and ZMSN@LEVO) were deposited in the sample chamber while the analysis one was filled with 3.1 mL of PBS. The system was kept at 37 ºC in an orbital shaker (100 rpm) refreshing the medium for each time of liquid withdrawing. A calibration curve for the LEVO was also acquired 0.02 to 20 µg/mL concentrations.

*6. Protein adhesion "in vitro" experiments*

*In vitro* adhesion of proteins was studied using lysozyme from egg white (90% Lys, L6876, Sigma-Aldrich), bovine serum albumin (96% BSA, A2153 Sigma-Aldrich), fibrinogen (75% Fib, F8630 Sigma-Aldrich) and protein-rich bovine fetal calf serum (FCS). A 50% v/v solution of the proteins (2 mg/mL for Lys, BSA and Fib, 20% v/v in the case of FCS in PBS 1x) was gently mixed with MSNs, ZMSNs and PEGylated MSNs dispersions (2 mg/mL in PBS 1x) and kept under orbital agitation (200 rpm) for 24h at 37 ºC. After protein exposure nanoparticles were centrifuged (10000 rpm) and washed with PBS in order to remove free or loosely bound proteins and a one-dimensional sodium dodecyl sulfate-polyacrylamide gel electrophoresis gel (SDS-PAGE) assay was performed. Briefly, the samples were mixed with a buffer (Tris 6 mM, SDS 2%, glycerol 10%, 2-β-mercaptoethanol 0.5 M, traces of bromophenol blue, pH = 6.8) and loaded in 10% SDS-PAGE gels. A calibration curve was estimated by loading 0.5, 1.0, 1.5, 2.0 and 2.5 µg/mL protein concentrations. The gels were run with a constant 100 V voltage (45-60 min) and stained in R-250 colloidal Coomasie blue solution for visualization.

*7. Cell uptake assay*

Macrophage-type murine RAW 264.7 were seeded ($5 \times 10^5$ cells/mL) on sterile 24-well plates and co-incubated for 90 min with 5 µg/mL suspensions of MSNs, ZMSNs and PEG-MSNs samples in complemented Dulbecco´s Modified Eagle´s Medium (DMEM) with 10% FBS at 37 °C with 5% $CO_2$. After this period cells were washed twice with PBS, and prepared for laser scanning confocal (LSCM) imaging. Therefore, cells were fixed with 4% (w/v) paraformaldehyde in PBS with 1% (w/v) sucrose (37 °C, 20 min), washed with PBS and permeabilized with buffered 0.5% Triton X-100 (10.3 g of sucrose, 0.292 g of NaCl, 0.06 g of $MgCl_2$, 0.476 g of 4-(2-hydroxyethyl)-1-piperazineethanesulfonic acid (HEPES), and 0.5 mL of Triton X-100 in 100 mL of deionized water, pH = 7.4) at 37°C. Non-specific binding sites were blocked with 1% (w/v) BSA in PBS (20 min at 37 °C). Actin filaments and nuclei were stained with Atto 565-conjugated phalloidin (dilution 1:40) and 40-6 diamino-20-phenylindole in PBS (1M, DAPI), respectively. LSCM images were taken with an Olympus FV1200 using 488 nm, 563 nm and 405 nm excitation wavelengths for nanoparticles, actin and nuclei, respectively. Alternatively, phagocytosis of the fluorescently labelled nanoparticles by RAW 264.7 macrophages was evaluated with a FACScalibur Becton Dickinson flow cytometer. After 90 min of incubation cells were trypsinized and an aliquot (500 µL) was taken and centrifuged (1400 rpm, 5 min) to recover cells. The obtained pellet was resuspended in PBS (300 µL) and trypan blue (20 µL) was added to quench fluorescence from surface bound nanoparticles, hence distinguish from internalized.

*8. Cell viability assay*

Proliferation of macrophages upon exposure to nanoparticles was determined based on the reduction potential of metabolically active cells obtained with AlamarBlue. Macrophages were seeded ($1x10^6$ cells/mL) on sterile 24-well plates and incubated with the MSN, ZMSN and PEGylated nanoparticles for 90 min at 37 °C. Afterwards 10 µL/well of a 4% AlamarBlue solution (v/v) in DMEM was added and incubated for 3h at 37 °C. Fluorescence was measured on a plate reader (Tecan Infinite F200, with $\lambda_{ex} = 560$ nm and $\lambda_{em} = 590$ nm). Measurements were taken at least two times in triplicate wells per condition in order to obtain proper statistics.

*9. Cytotoxicity of nanoparticles*

Biocompatibility and cytotoxic effects of the studied systems was tested via lactate dehydrogenase (LDH) assay in which the cellular membrane damage is related to release of cytosolic LDH enzyme (Spinreact S.A., Spain). The concentration of LDH is calculated considering its catalytic activity in the reduction of pyruvate by NADH which the converts non-fluorescent resazurin to fluorescent resorufin. The supernatant of nanoparticles-RAW 264.7 cells incubation experiments is taken and mixed with the LDH kit reagents (imidazole 65 mmol/L, pyruvate 0.6 mmol/L and NADH 0.18 mmol/L). Absorbance at 340 nm wavelength are measured at 0, 1, 2, 3 and 4 min, calculating the amount of enzyme that transforms 1 µmol of substrate per minute in standard conditions (U/L) as ΔA/min x 4925 (where ΔA is the average difference per minute and 4925 corresponds to the correction factor for 25-30 °C).

*10. Antimicrobial activity assays*

Effects against infection of the synthesized nanoparticles was evaluated by incubation with anaerobic common pathogens Gram negative *Escherichia coli* (*E. coli*) and Gram positive *Staphylococcus aureus* (*S. aureus*). Dispersions of unloaded (MSN, ZMSN) and drug loaded (MSN@LEVO and ZMSN@LEVO) nanoparticles were prepared at different concentration (5 and 10 µg/mL) in buffer solution. 500 µL of the suspensions were deposited on sterile 24- well plates containing 500 µL of highly concentrated bacterial medium (2. $10^8$ bacteria/mL, adjusted by $OD_{600}$) and cultured for 6 h in an orbital shaker (100 rpm) at 37 ºC. After this, aliquots were taken and subsequent dilutions (1/100, 1/1000, 1/10000) were performed prior to plate counting agar (PCA) evaluation of bacterial survival rate [33]. For this purpose, colony forming units (CFU) were calculated after seeding on Luria-Bertani (LB) agar plates and incubating 24 h at 37 ºC. Results of three independent experiments taken in triplicate were obtained through visual inspection and image analysis using adequate thresholding and size cut-off on FiJi software [38,39].

*11. Statistical Methods*

Experimental data are expressed as mean value ± standard deviation of a representative of three independent experiments carried out in triplicate. Statistical analysis of the significance of differences was done using one-way ANOVA analysis of variance. A value of $p <0.05$ was considered as statistically significant and denoted with an asterisk (*) in the figures. Statistical differences of $p <0.01$ were marked with two asterisks (**).

**Results and Discussion**

1. *Optimization of the synthesis of mixed-charge pseudo-zwitterionic MSNs (ZMSNs)*

MCM-41 silica nanoparticles (MSNs) are prepared via modified Stöber method [36,40]. FITC is incorporated to the MSNs as a tracer for cell internalization experiments by co-condensation reaction. Chemical modification of the labelled MSNs is constrained to take place on the outer surface of the nanoparticles, thus inner space for antibiotic loading is kept filled with the organic template [41] during the bifunctionalization processes. Covalent attachment of the amino and phosphonate-based molecules with the silanol groups of the silica surface (*c.a* 9.6 Si-OH/nm$^2$) [42,43] takes place through condensation reactions in a 1:3 stoichiometry. Fig. 1 displays the synthesis and functionalization processes to obtain ZMSNs materials.

The incorporation of the silanes through organic content of MSNs was quantified from TGA and elemental analysis (Table 1 and Fig. S1). Results show a variation of the theoretical to experimental APST:THPMSP ratio, evidencing that the amount of incorporated APST in the ZMSNs is, in all cases, superior to the theoretical one. This is explained by both the lower steric hindrance of the amino-silane molecule and the electrostatic repulsion between the negatively charged phosphonate groups and the silica matrix. .

In order to assess the functionalization process, changes in the ζ-potential values of the different samples were determined (Table 1). Pure silica MSNs ($pk_a$ = 6.8) exhibit a value of -30 mV due to of presence of deprotonated silanol groups (–Si-O$^{\ominus}$) onto the surfaces in physiological conditions. From the ζ-potential it is possible to establish an optimal pseudo-*zwitterionic* activity for a APST:THSMP theoretical molar ratio of 1:1.50 (standing for ZMSN-1.5). These functionalization composition yields a potential of (5 ± 3) mV, close to a global zero charge given by the existence of compensated NH$_3^{\oplus}$ and PO$_3^{\ominus}$ nuclei in an experimental 1:0.71 ratio. Regardless the prevalence of positive protonated amino groups provided by an excess of APST, we assume the almost null potential is reached due to the co-existence with unreacted surface –Si-O$^{\ominus}$ groups. The acid-base equilibrium and corresponding p$K_a$ of the different functional groups in aqueous medium are represented as follows:

$$R\text{-Si-OH} + H_2O \leftrightarrows R\text{-SiO}^- + H_3O^+ \qquad pK_a \approx 6.8 \qquad \text{Eq. (1)}$$

$$R\text{-PO}_3\text{-H} + H_2O \leftrightarrows R\text{-PO}_3^- + H_3O^+ \qquad pK_a \approx 2.0 \qquad \text{Eq. (2)}$$

$$R\text{-NH}_2 + H_2O \leftrightarrows R\text{-NH}_3^+ + OH^- \qquad pK_a \approx 10. \qquad \text{Eq. (3)}$$

**Table 1**: Characterization of the samples as a function of APTS:THSMP molar ratio obtained by: [a]elemental chemical analysis (atomic percentages), [b]DLS (hydrodynamic diameter, $D_H$) and [c]ζ-potential.

| Sample | APST:THSMP$_{molar}$ | | [a]%C | [a]%N | [b]$D_H$ (nm) | [c]ζ-potential (mV) |
|---|---|---|---|---|---|---|
| | Theoretical | Experimental | | | | |
| **MSN** | __ | __ | 5.68 | 0.15 | 190 ± 4 | -30 ± 2 |
| **ZMSN-1.00** | 1:1.00 | 1:0.18 | 9.99 | 2.59 | 164 ± 3 | 17 ± 2 |
| **ZMSN-1.25** | 1:1.37 | 1:0.37 | 10.27 | 2.43 | 122 ± 2 | 25 ± 2 |
| **ZMSN-1.44** | 1:1.44 | 1:0.40 | 12.55 | 2.53 | 164 ± 3 | 25 ± 2 |
| **ZMSN-1.50** | 1:1.50 | 1:0.71 | 12.19 | 1.80 | 220 ± 3 | 5 ± 3 |

The hydrodynamic diameter ($D_H$) of the synthesized nanoparticles is allocated in the range of 122-190 nm, which falls within the tolerance level for nanoparticles [43] and reflects no formation of large aggregates in aqueous media. In addition, the small value of their standard deviation reflex the homogeneity of the samples.

Fourier transform infrared spectra (FTIR) of pristine and functionalized MSNs (lowest (1:1.0) and highest (1:1.5) theoretical APST:THSMP) is shown in Fig. 2A. The existence of characteristics bands corresponding to Si-O-Si stretching (1100-900 cm$^{-1}$) and Si-O bending (469 cm$^{-1}$) as well as surface unreacted Si-OH (960 cm$^{-1}$) confirm that the ordered dense silica network is maintained after functionalization [35]. Besides these, it is observed a peak at 560 cm$^{-1}$ which determines defects from the crystalline structure [44]. The broad band in the range from 3470 to 3450 cm$^{-1}$ is attributed to the O-H stretching vibration of water molecules either adsorbed to the surface Si-OH moieties or to themselves through hydrogen bonding [45]. Furthermore, the existence of unreacted surface Si-OH groups is revealed from the band located at 960 cm$^{-1}$. The grafting reaction of the positively charged molecules (APST) introduces new bands corresponding to stretching and deformation modes of protonated $NH_3^{\oplus}$ groups at 3550 and 1590 cm$^{-1}$, respectively, as well as low intensity primary NH stretching and deformation frequency peaks located at 3550 and 1643 cm$^{-1}$, respectively [31]. Bands at 2952 and 2880 cm$^{-1}$ correspond to symmetric and asymmetric stretching vibrations of unreacted propyl residues from the APST. The negatively charged part of the mixed-charge surface (phosphonate) appears in the range from 1215 cm$^{-1}$ to 946 cm$^{-1}$. As shown in the inset of Fig. 2B, the low intensity peak at 1215 cm$^{-1}$ is assigned to the –P=O linkage of the THSMP [46], whereas the peak located around 975 cm$^{-1}$ corresponds to the –P-O(H) linkage from the R-$PO_3^{\ominus}$ [47]. At lower wavelengths stretching bands from P-P and P-OH are found (1007 and 946 cm$^{-1}$, respectively). Therefore, the existence of $NH_3^{\oplus}$ and $PO_3^{\ominus}$ chains confirm the pseudo-*zwitterionic* modification of MSNs. The optimal aminopropyl:phosphonate molar ratio is established to be theoretical 1:1.5, thus results will be summarized to ZMSN-1.5 material henceforth. Pseudo-*zwitterionic* behavior is furthermore confirmed through zeta-potential variation as a function of pH. Results show a pH-dependent ζ-potential behavior (Fig. S5) given by the contribution of the amino (overall positive charge) or the combined phosphonate and silanol moieties (overall negative values) from acidic to basic conditions, respectively. A *zwitterionic*-like charge distribution of overall zero potential is observed for the physiological pH range (6.9 to 7.5).

XRD patterns (Fig. 3A) reveal a 2D hexagonal mesoporous arrangement for both bare MSN and functionalized ZMSN-1.5 given by the indexed 10, 11, 20 and 21 reflections characteristics of the *p6mm* crystallographic group [11]. TEM images confirm the presence of spherical nanoparticles with a honeycomb-like mesoporous structure corresponding to 2D hexagonal arrangement for both bare MSN and functionalized ZMSN-1.5 (inset, Fig. 3A). Variations in the textural features of the nanoparticles upon pseudo-*zwitterionization* are calculated from nitrogen adsorption/desorption porosimetry and summarized in Table 2. Both isotherms are IV type according to the IUPAC, which are present in mesoporous materials with parallel cylindrical pores in a 2D hexagonal structure [48]. Regarding to $S_{BET}$ and $V_P$, they display a decrease in ZMSNs compared to pristine MSNs, because of the post-grafting of organic moieties [49]. This decrease could be attributed to the small size of both functionalizing molecules, which may partially block the pore entrances, as previously has been reported [13]. These findings are in good agreement with the calculated pore diameter ($D_P$) and wall thickness ($t_{wall}$) values, which remain constant in ZMSNs, because the functionalization takes place in the outermost silica surface. Nevertheless, the textural parameters of ZMSN-1.5 draw appropriate host characteristics for drug loading and targeting delivery on the cylindrical pores (parallel arrangement) of the synthesized mesoporous materials.

**Table 2.** Textural features ($S_{BET}$: surface area; $V_P$: pore volume; $D_P$: pore diameter; $t_{wall}$: wall thickness; $a_0$: lattice parameter) of pristine MSN and mixed-charge functionalized ZMSN-1.5.

| Sample | $S_{BET}$ (m²/g) | $V_P$ (cm³/g) | $D_P$ (nm) | $t_{wall}$ (nm) | $a_0$ (nm) |
|---|---|---|---|---|---|
| MSN | 1041 | 1.20 | 2.40 | 2.10 | 4.60 |
| ZMSN-1.5 | 500 | 0.48 | 2.30 | 2.40 | 4.70 |

Solid state NMR spectroscopy is used to assess the covalent bi-functionalization of the MSNs (Fig. 4). In the same way, the PEGylated MSNs have been also characterized and the results are collected in the supporting information (Fig . S3). From the $^1H \rightarrow {}^{13}C$ CP-MAS NMR spectra of the nanoparticles (Fig. 4A) it is observed the presence of residues from the templating surfactant and the presence of protonated amino-silane (APST) and deprotonated phosphonate-silane (THSPMP) to the surface of the silica framework, confirming the polyampholyte nature on the ZMSN surface. Concerning $^{29}$Si MAS NMR studies (Fig. 4B), the appearance of $T^n$ bands (R-Si(OSi)$_n$(OH)$_{3-n}$, n = 2-3) arising at -59.6 ($T^2$) and -68.1 ppm ($T^3$) in the ZMSN-1.5 shows the presence of silane atoms bound to an alkyl group and the reaction of Si-OH groups of silane derivatives with other Si-OH groups, but it cannot clearly distinguish if the latter ones belong to the silica surfaces, to other silane molecules or to a combination of them. In other hand, the $Q^n$ region resonances located at ca. -93 ($Q^2$), -101 ($Q^3$) and -111 ppm ($Q^4$) chemical shifts correspond to $Q^n$ = Si(OSi)$_n$(OH)$_{4-n}$, n = 2-4, where n values stand for geminal silanol (n = 2), one free silanol (n = 3) or fully condensed silica backbone ($Q^4$). In this sense, the conversion of Si-OH groups of bare MSNs to fully condensed Si-O-Si species after the post-grafting reactions is reflected in the decrease on the $Q^2$ and $Q^3$ peak areas and an increase in $Q^4$ peak area, i.e. a decrease in the ($Q^2$ + $Q^3$)/$Q^4$ ratio (Table 3). This fact confirms the existence of covalent linkages between silica surfaces and the organic groups in the bifunctionalized samples as previously has been reported for other silica materials [10]. In addition, the 31P MAS NMR profile verifies the existence of phosphonate species on the surface of MSN with the sharp and intense peak located at 24.2 ppm. Therefore, probably a combination of a silsesquioxane network around MSNs together with covalent linkages with the silanol groups from the silica surface is occurring in these samples.

**Table 3.** Peak area (%) of deconvoluted silicon $Q_n$ and $T_n$ $^{29}$Si MAS NMR bands.

| Sample | $T_2$ | $T_3$ | $Q_2$ | $Q_3$ | $Q_4$ | $(Q_2+Q_3)/Q_4$ |
|---|---|---|---|---|---|---|
| MSN | — | — | 8.7 | 48.3 | 42.9 | 1.33 |
| ZMSN-1.5 | 7.0 | 11.5 | 5.5 | 24.0 | 52.1 | 0.56 |

2. *Protein adsorption*

Due to the need of relatively long bloodstream circulation times to reach the target and considering the tendency of these nanocarriers to undergo nonspecific protein adsorption when encounter biological fluids, *in vitro* adsorption of several proteins (lysine, bovine serum albumin, fibrinogen and fetal calf serum) are tested. The low-adsorbed proteins on the surface of these nanosystems will determine its success in an *in vivo* scenario. Fig.S4 shows results from electrophoresis with buffered Laemmli gel (pH = 7.4, 37 °C). ZMSN-1.5 exhibit a significantly lower adsorption against low size proteins such as Lys (14.4 kDa), BSA (67 kDa) and Fib (340 kDa) compared to pristine MSNs. Furthermore, the reduction in protein corona is comparable to that one yielded by the typical PEGylated nanoparticles. The protein adsorption is estimated to be reduced in an 80-90% upon functionalization. Considering these results, it is expected that our fabricated nanocarriers will not undergo clearance by the immune system as no protein corona is formed, thus can effectively deliver the required antibiotic to the target infection [16].

*3. In vitro cell internalization*

As it has been previously proposed, cellular uptake of nanoparticles and distribution within the cell body is a process governed by several parameters including size, shape, charge and chemistry [50,51]. In our study we evaluated internalization into RAW 264.7 murine macrophague cells of our systems firstly by laser scanning confocal microscopy upon incubation with 5 µg/mL nanoparticles dispersions in culture medium for 90 min (Fig. 5A). Through labelling cell nuclei, nanoparticles and actin filaments with molecules of varying excitation/emission wavelengths (DAPI, $\lambda_{exc}$ = 405 nm, FITC, $\lambda_{exc}$ = 488 nm, Atto-565-phall, $\lambda_{exc}$ = 563 nm, respectively), it is possible to run co-localization analysis of the nanocarriers. Efficient uptake by macrophagues takes place, as expected, when placing in contact with pristine MSNs. Silica nanoparticles are localized in the cell cytoplasm not entering the cell nuclei. Cell uptake is drastically reduced for functionalized nanoparticles, both ZMSN-1.5 and PEGylated. In both cases, fluorescence arising from the green labelled nanoparticles is mainly localized on the border of the cell membrane, thus the particles are adsorbed and do not undergo engulfment through vesicle formation during the studied 90 min incubation period. The ratio of total average fluorescence intensity arising from the nanoparticles (green channel) between those within the cells is calculated by image analysis with FiJi software (Table SI). Naked MSNs are undergo a 30-40% uptake, while bi-functionalization with mixed-charge brushes and PEG reduces engulfment up to approximately 17 and 8%, respectively. This estimation is confirmed by flow cytometry (Fig. 5C-D), where a reduction of approximately 37% of internalization it is calculated for ZMSN-1.5 compared to bare MSN. Therefore, the pseudo-*zwitterionic* modification of the MSNs provides proper stealth properties against immune system since a reduced macrophage cell uptake is observed. Furthermore cytotoxic effects on RAW 264.7 cells were evaluated through LDH and Alamar Blue tests (Fig. 5B). A slight decrease in cell viability was appreciated for all the studied nanoparticles in contact with cells compared to control fluorescence, although an overall survival ca. 80% was achieved. In the case of LDH experiment, bare MSNs showed a certain cytotoxic effect on macrophages, which was noticeably reduced for the PEGylated and mixed-charge pseudo-*zwitterionic* MSNs systems. These findings evidence that since both parameters (low-fouling and reduced cell uptake) are comparable in functionalized samples (ZMSN-1.5 and PEG-MSNs) and significantly lower with respect to bare MSNs, the pseudo-*zwitterionic* surfaces could constitute an alternative to PEGylation. In this sense, such surfaces could overcome the oxidative process and formation of undesirable subproducts of PEGylated materials [15,21].

*4. "In vial" antibiotic (LEVO) release*

In order to establish the antibiotic releasing efficiency of the low-fouling ZMSN-1.5 nanoparticles under physiological conditions (pH = 7.4, PBS 1x), we evaluated the time evolution of the total amount of antibiotic released normalized with the initial amount of loaded molecule ($W_T = W_0$) through impregnation method (Fig. 6). This amount of loaded LEVO was estimated through elemental analysis at approximately 2% for MSNs and 4% for ZMSN-1.5. These results could be explained based on a greater affinity of the amine entities present in ZMSNs towards LEVO moieties, which favors the incorporation of the antibiotic inside the mesopores, as is it has also been described in amine-functionalized MSNs materials [13,52]. Furthermore, $N_2$ adsorption porosimetry data of loaded systems verified a much larger reduction in surface area, pore size and volume for the ZMSN-1.5 systems, which is in accordance with the larger extent of levofloxacin insertion within the mesopores (see Supporting Information, Table S1). The theoretical kinetic model for these systems can be adjusted to a first order kinetic model with the Chapman equation (4):

$$W_T/W_0 = A (1-e^{-\lambda t})^\delta \qquad \text{Eq. (4)}$$

where $W_T/W_0$ represents the percentage of LEVO released at a given time (t), A is the maximum amount of drug released to the medium, $\delta$ represents a non-ideality parameter, (included as an indicator of the deviation from linearity) and $\lambda$ the release rate constant [13]. ZMSN-1.5 performs a faster drug release ($\lambda = 0.14$ h$^{-1}$) than pristine MSNs, in a first-order kinetics trend ($\delta = 1$) with and established effective release of approximately 40%, and lowering the velocity after 25h. On the other hand, despite the kinetics of MSNs is slower at short times, its evolution shows a smoother slow down with a possible continuous release even after 60 h in contact with the medium. As stated by González *et al* in the study of dendrimer decorated MSNs, bare silica nanoparticles are capable to strongly retain LEVO molecules (up to 50% within 14 days) due to hydrogen bonding [13,53]. On the contrary, interactions of the antibiotic molecule with protonated amino and phosphonate groups of the ZMSN aid a faster liberation of LEVO to the medium. Despite a rapid releasing time for ZMSN compared to bare MSNs, both samples are capable to deliver comparable amounts of antibiotic at the studied times, therefore pseudo-*zwitterionization* through bi-functionalization does not affect the amount of delivered drug.

## 5. Antimicrobial effect

After evaluating the low-fouling and furtive properties and antibiotic release capability of these mixed-charge pseudo-*zwitterionic* nanosytems *in vitro* antimicrobial assays against pathogenic *E. coli* and *S. aureus* were performed to assess the efficacy of these nanosystems in an infection scenario. Fig. 7 shows the antimicrobial effect of pristine MSN and ZMSN with and without LEVO at lowest concentration (5 µg/mL) by measurement of the colony forming unit (CFU)/mL. Pseudo-*zwitterionic* functionalization shows an averaged reduction in bacterial growth allocated *circa* 47% and 74% for *E. coli* and *S. aureus*, respectively, which is in accordance with the inherent antimicrobial effect of phosphonate group of the mixed-charge brushes [54]. On the other hand, upon loading with LEVO (5@ and 10@ for the tested nanoparticle suspension concentrations), a notable decrease in bacterial survival is shown in both matrices (MSNs and ZMSNs) due to the releasing of levofloxacin to the media. This effect is dosage-dependent, showing high antimicrobial effectiveness even to low doses (5 µg/mL). However, significant differences between both samples containing LEVO are observed, showing higher antimicrobial effectively for ZMSN samples compared to MSNs. This is due to a synergistic effect between the anionic functionalizing agent (THSPMP) with inherent antimicrobial properties [35] and the antibiotic release loaded in the mesoporous structure. In this sense, an antimicrobial efficacy of 99.9 and 100% for ZMSN@LEVO are observed for Gram negative and positive bacteria, respectively. Note that for MSN@LEVO, the histogram displays bacterial survival above $10^5$ CFU after 6 h of incubation in both pathogens, which can be considers as inefficient. It has been reported that a few live bacteria as 10–100 CFU/mL can cause an infection, increasing the bacterial resistant [55]. However, the CFU/mL analysis for ZMSN@LEVO nanosystems display a remarkable efficiency against both Gram-negative and Gram positive bacteria, showing values below of this values with significant efficacy.

**Conclusions**

Pseudo-*zwitterionic* antimicrobial MSNs with low-fouling and reduced macrophage-uptake behavior has been developed. The pseudo-*zwitterionic* behavior has been acquired by a simple method consisting in simultaneous direct grafting of short chains of aminopropyl silanetriol and trihydroxy-silyl-propyl-methyl-phosphonate mixed-charge brushes onto the external MSN surface. This nanocarrier has been tested for local infection treatment through the synergistic effect of levofloxacin loaded into the mesoporous structure of the nanosystems and the special release profile given by the drug-to-charges electrostatic interactions. An excellent antimicrobial activity (>99.9%) against pathogenic *E. coli* and *S. aureus* has been observed even for low nanoparticles concentration (5 μg/mL). The development of this simple mixed-charge pseudo-*zwitterionic* MSN based antimicrobial nanocarriers is envisioned to effectively contribute to the field of bacterial infection.

**Acknowledgments**

This work was supported by European Research Council, ERC-2015-AdG (VERDI), Proposal No. 694160; Ministerio de Economía y Competitividad (MINECO) grants MAT2015-64831-R, MAT2016-75611-R AEI/FEDER and Juan de la Cierva Incorporación (IJCI-2015-25421) fellowship (N. E.). The authors wish to thank the ICTS Centro Nacional de Microscopia Electrónica (Spain), CAI X-ray Diffraction, CAI elemental analyses, CAI NMR, CAI Cytometer and Fluorescence microscopy of the Universidad Complutense de Madrid (Spain) for the assistance.

**Figure and Figure captions**

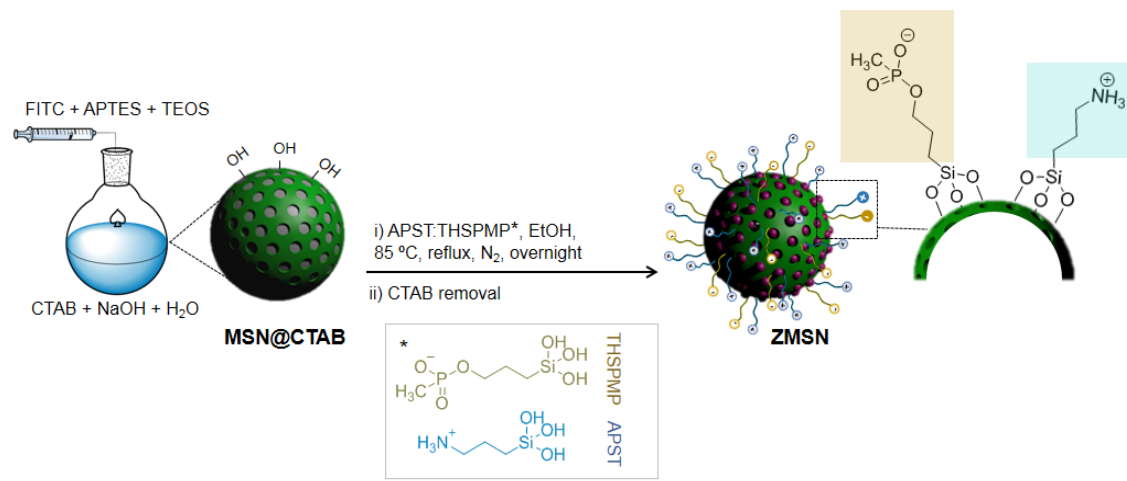

**Fig. 1.** Synthesis of labelled mixed-charge pseudo-*zwitterionic* MCM-41 nanoparticles (ZMSNs)

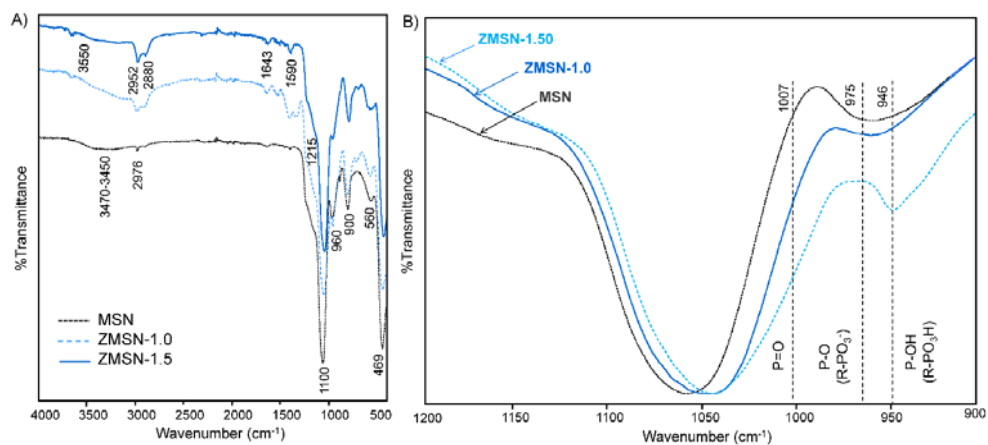

**Fig. 2.** A) ATR-FTIR spectra of pristine MSN and functionalized ZMSN-1.0 and ZMSN-1.5. B) Zoom of the phosphonate bands frequencies

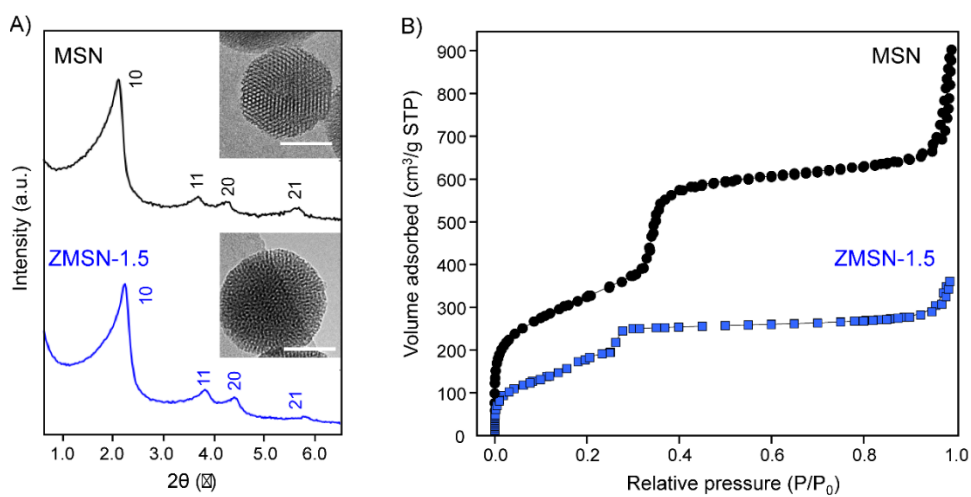

**Fig. 3.** A) Low angle X-ray diffraction patterns and B) $N_2$ adsorption-desorption isotherms of the i) MSN and ii) ZMSN-1.5 nanoparticles. Insets: TEM images (*Scale bar: 50 nm*).

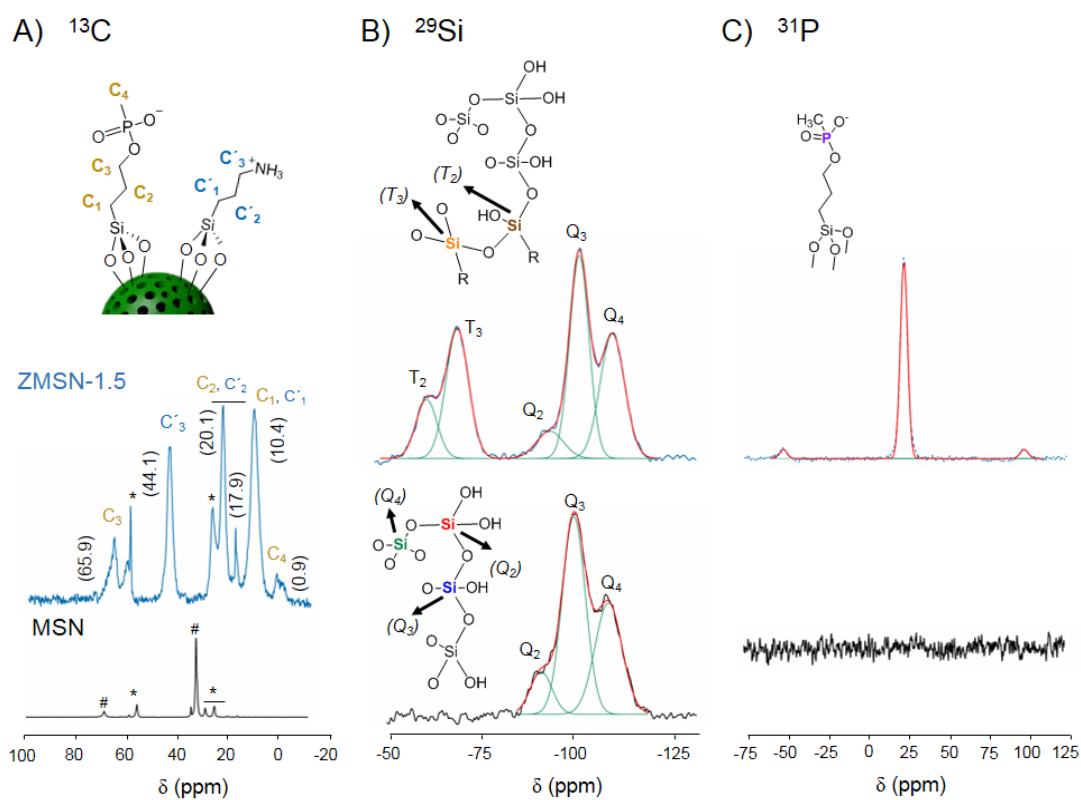

**Fi. 4.** A) $^1H \rightarrow {}^{13}C$ CP-MAS NMR spectra of MSN and ZMSN-1.5 nanoparticles, showing the carbon assignment in the upper structure scheme. Peaks corresponding to residual CTAB and ethoxy carbons from incomplete hydrolysis and condensation reactions are marked as well with # and *, respectively. B) Cross-polarized $^{29}Si$ MAS NMR spectra of MSN and ZMSN-1.5. Positions and correspondence of the $Q_n$ and $T_n$ bands within the silica network are marked. C) $^{31}P$ MAS NMR spectra of pristine MSN and ZMSN-1.5, showing the peak corresponding to phosphonate species.

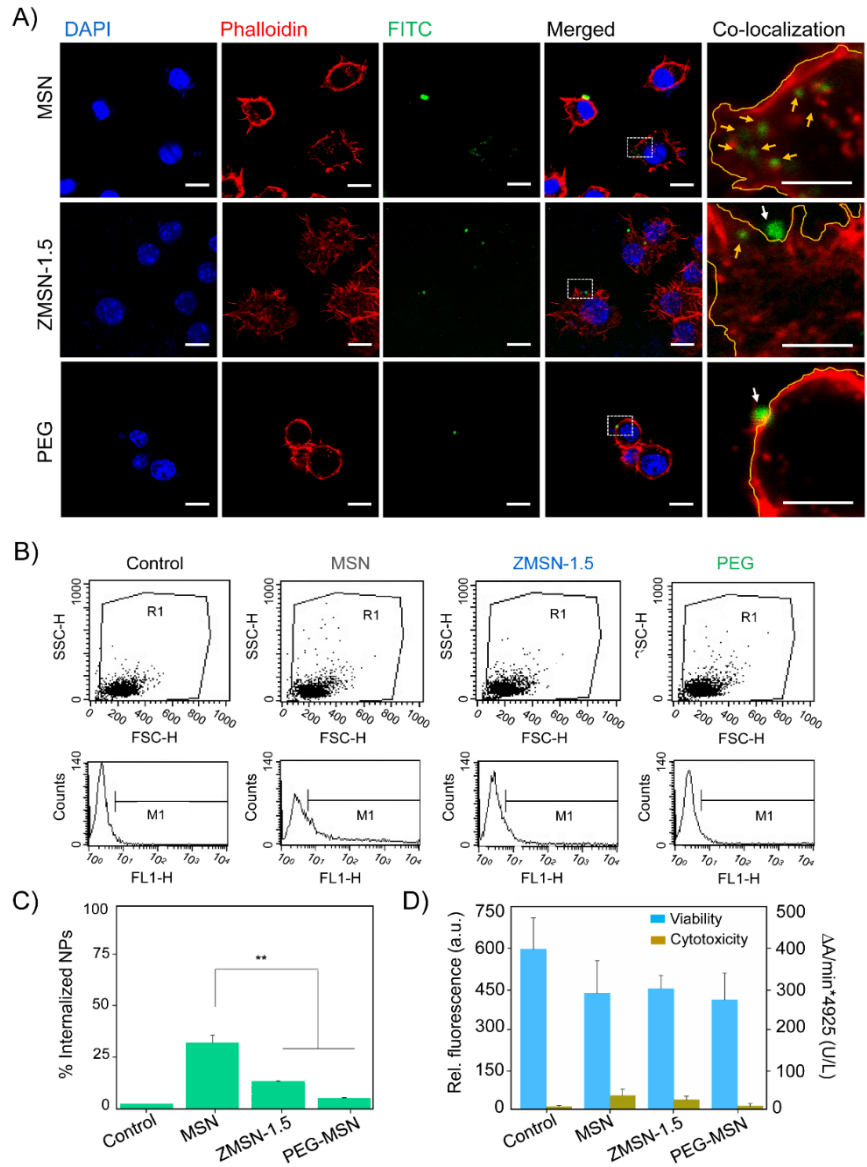

**Fig. 5.** A) Laser scanning confocal images of the nuclei (DAPI), membrane (Phalloidin) and nanoparticles (FITC) emission channels. Merged images and high magnification merged red and green channels overlay allow to co-localize the different studied systems (bare MSNs, pseudo-*zwitterion* ZMSN-1.5 and control PEGylated MSNs). In the co-localization right row area selection of region of interest was done with FiJi, marking in yellow the cell membrane border. Internalized nanoparticles are highlighted with yellow arrows, while those located in the outer area are marked with white arrows. (*Scale bar: 10 μm, 5 μm for co-localization row*). B) RAW 264.7 macrophague cell survival rate (left axis, blue) and cytotoxic effect of the nanoparticles (right axis, yellow) after 90 min incubation at 37 °C. Experiments are performed using Alamar Blue and LDH kits, respectively. C) Flow cytometry side (SSC-H) versus forward (FSC-H) scatter plots (*upper row*) and count histograms (*lower row*) showing the internalization of FITC-labelled nanoparticles. D) Statistical internalization analysis for the different systems given mean percentage of duplicate experiments. Functionalized pseudo-*zwitterion* and PEGylated MSNs are compared with naked particles, **$p < 0.01$.

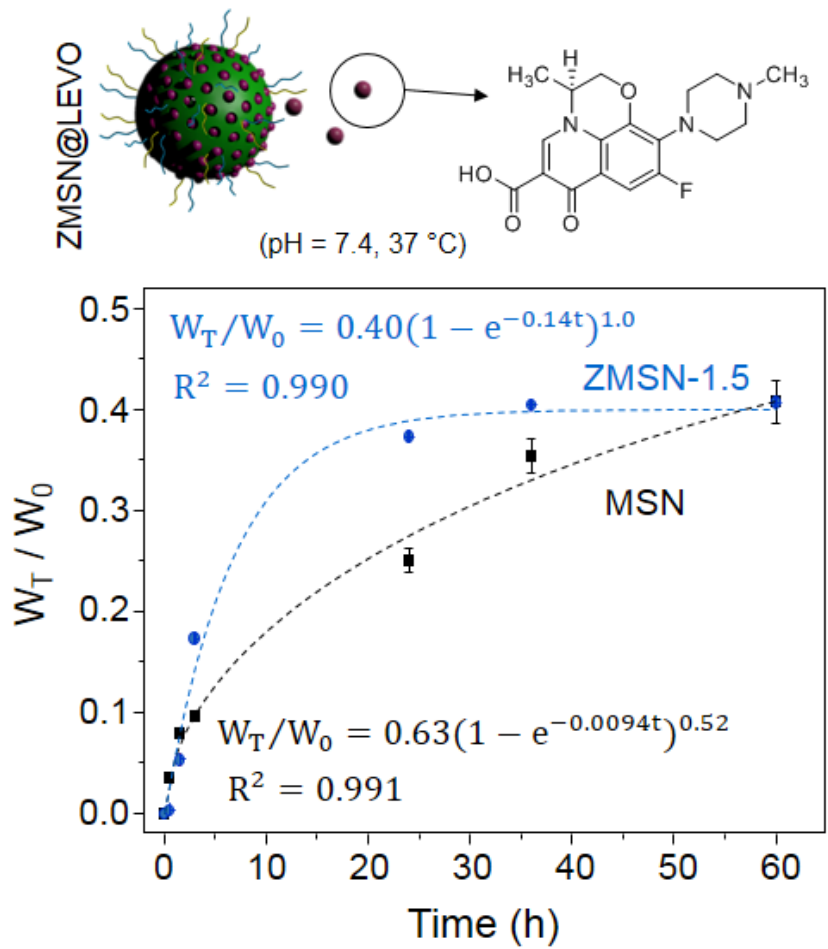

**Fig. 6.** *In vitro* cumulative LEVO release kinetics of bare MSN and mixed-charge pseudo-*zwitterionic* nanoparticles showing the structure of the antibiotic in the sketch.

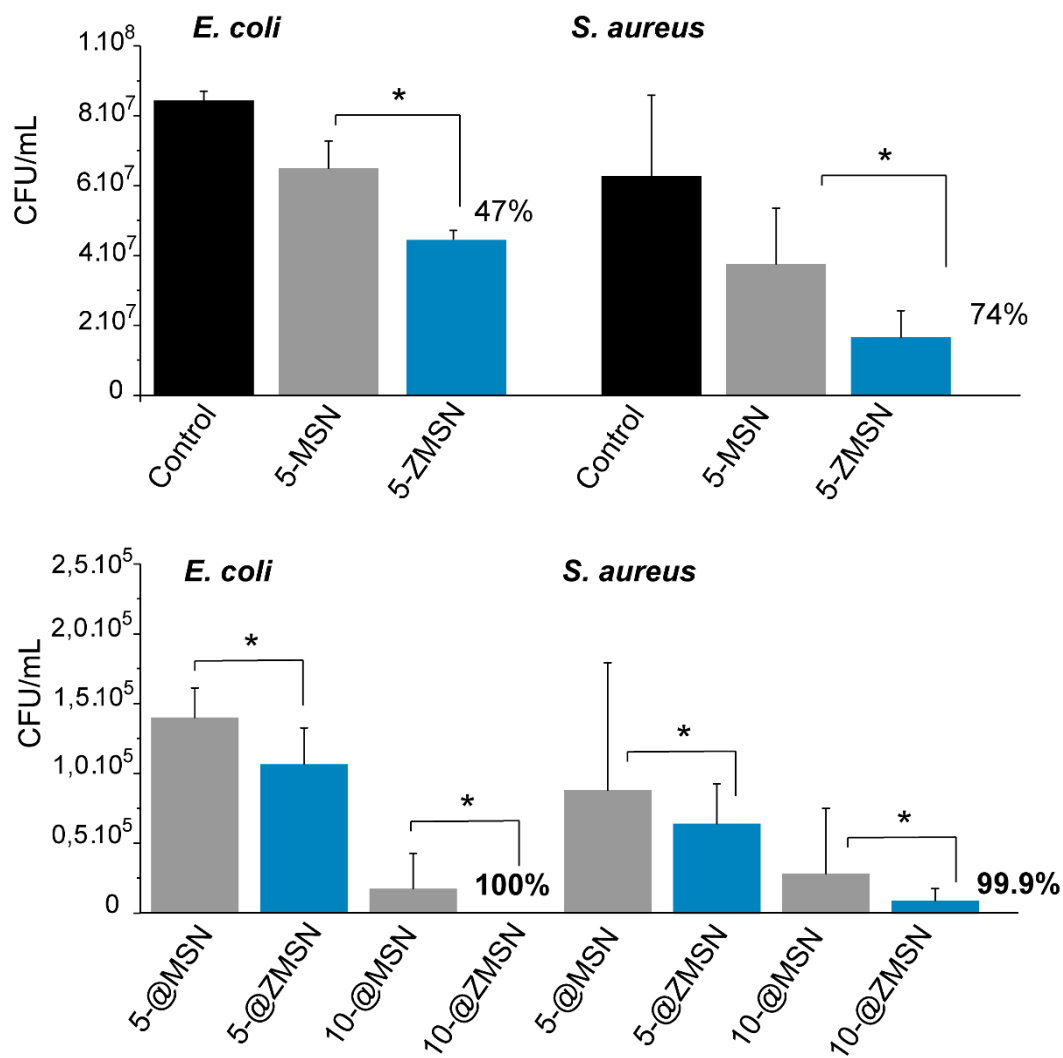

**Fig. 7.** Antimicrobial activity of bare (MSN) and functionalized (ZMSN-1.5) nanoparticles both unloaded (Top set) and LEVO loaded (@,at different concentration ) (Bottom set) against gram-negative *E. coli* and gram-positive *S. aureus*. Results are expressed as colony forming unit (CFU) per mL growth of microorganism after 6h incubation. Unloaded samples the lowest concentration is shown in order to elucidate the antimicrobial effect. Loaded samples different concentration are shown (5, 10) µg/mL in PBS in order to determine the dosage effectiveness. Results and standard deviations are calculated from triplicates of three independent experiments, $^*p < 0.05$.